# Biomimetic Polymer Film with Brilliant Brightness Using a One-Step Water Vapor–Induced Phase Separation Method

*Weizhi Zou, Lorenzo Pattelli, Jing Guo, Shijia Yang, Meng Yang, Ning Zhao\*, Jian Xu\*, Diederik S. Wiersma\**

W. Zou, J. Guo, S. Yang, M. Yang, Prof. N. Zhao, Prof. J. Xu
Beijing National Laboratory for Molecular Sciences, CAS Research/Education Center for Excellence in Molecular Sciences, Laboratory of Polymer Physics and Chemistry, Institute of Chemistry, Chinese Academy of Sciences, Beijing, 100190, P. R. China
E-mail: zhaoning@iccas.ac.cn; jxu@iccas.ac.cn

W. Zou, J. Guo, S. Yang, M. Yang, Prof. J. Xu
University of Chinese Academy of Sciences, 100049, Beijing, P. R. China

Dr. L. Pattelli, Prof. D. S. Wiersma
European Laboratory for Non-linear Spectroscopy (LENS), Università di Firenze, Sesto Fiorentino, (FI), Italy
E-mail: wiersma@lens.unifi.it

Dr. L. Pattelli, Prof. D. S. Wiersma
Istituto Nazionale di Ricerca Metrologica (INRiM), Turin, Italy

Prof. D. S. Wiersma
Department of Physics, Università di Firenze, Sesto Fiorentino, (FI), Italy

The scales of the white *Cyphochilus* beetles are endowed with unusual whiteness arising from the exceptional scattering efficiency of their disordered ultrastructure optimized through millions of years of evolution. Here, a simple, one-step method based on water-vapor induced phase separation (VIPS) is developed to prepare ultra-thin polystyrene (PS) films with similar microstructure and comparable optical performance. A typical biomimetic 3.5 μm PS film exhibits a diffuse reflectance of 61% at 500 nm, which translates into a transport mean free path below 1 μm. A complete optical characterization through Monte Carlo simulations reveals how such scattering performance arises from the scattering coefficient and scattering anisotropy, whose interplay provides insight into the morphological properties of the material. The potential of bright-white coatings as smart sensors or wearable devices is highlighted using a treated ultra-thin film as a real-time sensor for human exhalation.

Disordered materials exhibiting a bright, broadband diffuse reflection[1-3] play a key role in many applications, including passive-daytime-radiative-cooling (PDRC) materials,[4-6] displays,[7] LEDs[8-10] and light harvesting devices,[11,12] to name a few. In general, achieving a bright white appearance requires either a thick scattering layer or scattering elements with high refractive index $n$.[3,13,14] Indeed, to date, titanium dioxide (TiO$_2$, $n \approx 2.67$) nanoparticles represent the most widely used pigment for white coatings and paint.[15] Interestingly, though, the natural world offers several examples of white-appearing organisms featuring thin scattering layers despite not having access to highly scattering materials such as metal oxides or semiconductors. One prominent example is that of the beetles of the *Cyphochilus* genus,[3,16] which are covered by an array of thin scales (7±1.5 μm) made of a three-dimensional disordered network of chitin filaments which can diffusely reflect up to ~ 70% of incoming light over a broad wavelength spectrum (**Figure 1**a).[1,3,13,17-19] Such optical performance is all the more outstanding considering the relatively low refractive index of chitin (~1.56)[17,18] and the high volume fraction occupied by the network (~ 50%), with respect to the lower density typically expected for optimally scattering materials.[3,16,19,20] For this reason, the optimized network structure of this beetle has raised a lot of interest in different fields, as it suggests that highly efficient scattering materials can be obtained from more readily available low refractive index materials such as polymers. Even more, its network structure holds promise for the development of particle-free scattering media, which are being actively sought after in recent years due to health and environment related concerns about nanoparticle-based scattering media.[21-24]

To date, methods such as electro-spinning,[25] supercritical carbon dioxide foaming (SC-CO$_2$)[24] and cellulose nanocrystal (CNC) self-assembly[14] have been proposed to prepare bright-white surfaces using polymers with low refractive index, but have failed to match the optical performance of the *Cyphochilus* beetles. More recently, poly(methyl methacrylate) (PMMA) films prepared by a non-solvent induced polymer phase separation (NIPS) method[26] have been reported exhibiting a scattering strength comparable to that of the beetle, but the relatively complex preparation process of sonication-induced degradation to obtain PMMA with low polydispersity in order to control the phase separation process and form a network structure represents a limitation for large-scale preparation and practical applications. Moreover, despite increasing efforts to engineer novel materials inspired by the *Cyphochilus* beetles, the interplay between its scattering performance and the complex network topology is still not entirely understood, with several aspects proposed to explain its outstanding optical properties.[13,19,27] Additionally, comparisons between different scattering materials have been mostly limited so far to measurements of total integrated reflectance which, despite being the relevant quantity for most applications, fails to provide any insight on the single scattering properties of the material. In this respect, extending optical characterization to the single scattering level represents an important step forward to build a scattering "fingerprint" for the growing array of bioinspired materials, gaining insights on potentially different scattering mechanism underlying their whiteness.

In this work, we describe a one-step spin-coating assisted water-vapor induced phase separation (VIPS) method that does not involve any pre- or post-processing to fabricate biomimetic bright-white polystyrene (PS, $n \approx 1.59$)[28] films. Using this technique, we obtained highly efficiently scattering layers able to reflect up to 61% (subtracting specular reflection) at 500 nm wavelength for a free-standing film with a thickness of just 3.5 μm. We retrieved the scattering parameters of the resulting films by solving the radiative transfer equation (RTE) through a Monte Carlo look-up table approach[29] to obtain a full characterization of the scattering mechanism behind their optical performance. The simple preparation strategy and in-depth optical characterization provide useful information to design and fabricate novel network-based materials with advanced optical properties. In addition, we show how such thin film can be used as a real-time optical switch triggered by human breath exploiting its excellent brightness and activatable super-hydrophilicity. Owing to the good mechanical stability and flexibility of the material, we envision that bright-white coatings will play an increasingly important role in the emerging field of smart sensors and wearable devices for monitoring purposes.[30-32]

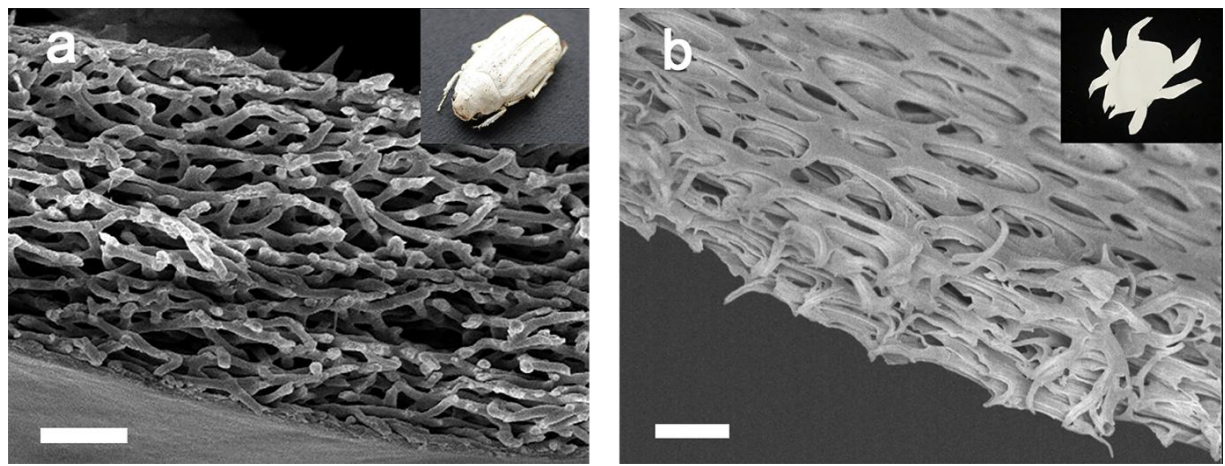

**Figure 1.** (a) SEM image of the cross-section of the *Cyphochilus* scale. Scale bar: 2 μm. Inset: photo of the *Cyphochilus* white beetle. (b) SEM image of the cross-section of the 3.5 μm PS film. Scale bar: 2 μm. Inset: photo of the free-standing 3.5 μm film cut out with a white beetle shape, showing its good mechanical stability.

A typical PS film prepared using the VIPS method is shown in Figure 1b. A 25 wt% PS solution in *N,N'*-dimethylformamide (DMF) was spin-coated in an environment with a relative humidity of 95%. PS films with various thicknesses can be obtained by controlling the spin-coating speed and the volume of polymer solution used. At a thickness of 3.5 μm, the film exhibits a typical disordered bi-continuous structure made by sub-μm backbone blocks with a polymer volume fraction of about 60% (See Experimental section). Morphologically, the bi-continuous structure exhibits a laminated arrangement as shown in the SEM image, highlighting the anisotropic structure of the polymer network between the in-plane and out-of-plane directions. The thin film shows a broadband diffuse reflectance (i.e., calculated excluding the specular component, see Experimental section) over the visible range, varying from 64% at 400 nm to 52% at 800 nm (**Figure 2**a).

We evaluate the scattering performance of the 3.5 μm PS film in terms of the scattering mean free path $\ell_s$ and the transport mean free path $\ell_t$, representing respectively the average distance between two consecutive scattering events and the average distance over which the original propagation direction is completely randomized. These two parameters are connected by the degree of scattering asymmetry parameter *g* through the so-called similarity relation $\ell_t=\ell_s/(1-g)$, where *g* is the average cosine of the scattering angular distribution and can typically vary from 0 (i.e. isotropic scattering) to 1 (i.e. forward propagation).[33] Although less common, it should be noted that the asymmetry factor can also become negative (resulting in $\ell_t<\ell_s$) in case of resonant scattering or strong positional correlations.[34,35]

In the diffusive regime, the total reflectance is mainly determined by the ratio between the slab thickness and the transport mean free path, according to the so-called Ohm's law for light.[36] Indeed, this simple approach has been used several times in the literatures, even though its validity breaks down at low sample thickness.[37,38] In our case, due to the reduced thickness of the films, a residual ballistic transmitted intensity can be collected with a distant optical fiber (see Experimental Section) allowing a complete optical characterization that does not rely on Ohm's law nor on the diffusive approximation. We solved the radiative transfer equation (RTE) through a Monte Carlo (MC) approach, enabling the retrieval of scattering parameters in configurations well beyond the validity range of diffusion theory (see Supporting Information). In particular, by assuming for simplicity a structurally isotropic medium, we retrieve the scattering parameters along the perpendicular direction to the film,[39] which are the ones determining the overall turbidity.[13] Even in this simplified picture, to our knowledge, we provide the first broadband characterization of the scattering asymmetry parameter *g* for network-like materials, providing novel information on their structural properties. The estimated optical parameters are obtained by fitting the experimental measurements of collimated (i.e., ballistic) transmission and integrated diffuse reflectance with Monte Carlo calculations, as shown in Figure 2b and 2c for the 3.5 μm film (see Supporting Information).

As can be seen, the retrieved transport mean free path along the perpendicular direction of the slab is comparable to that of the *Cyphochilus* beetle ($\ell_t$=1.47 μm)[3] at near infrared frequencies, and is appreciably shorter in the visible range, falling below 1 μm for $\lambda < 515$ nm. Interestingly, we find that the simple monotonic trend of $\ell_t$ results from a more complex interplay between the scattering anisotropy and the scattering mean free path as retrieved from ballistic transmission measurements (Figure 2b). Indeed, the scattering mean free path reaches a remarkably low minimum of 0.41 μm at a wavelength of 600 nm, which is however compensated by a peak in the scattering asymmetry. This interplay between scattering efficiency and asymmetry – and in particular their spectral dependence over visible wavelengths – allows to infer the dominant scattering mechanism determining the white appearance. As regards the spectral dependence of the parameter *g* (Figure 2c), there is a general lack of reports in the literature, especially for fiber-based or structurally anisotropic materials. A potentially related example is that of dentine, whose internal structure characterized by aligned tubules might also give rise to a peak of the scattering asymmetry in the visible range.[40,41] As described in the following, our results support this hypothesis and suggest that the scattering asymmetry peak position and amplitude can be used to infer about the size distribution of the elongated scattering elements (see Supporting Information). In this respect, the spectral dependence of the scattering anisotropy represents an untapped source of information that could be used to tell apart different scattering mechanisms in thin layers of turbid materials.

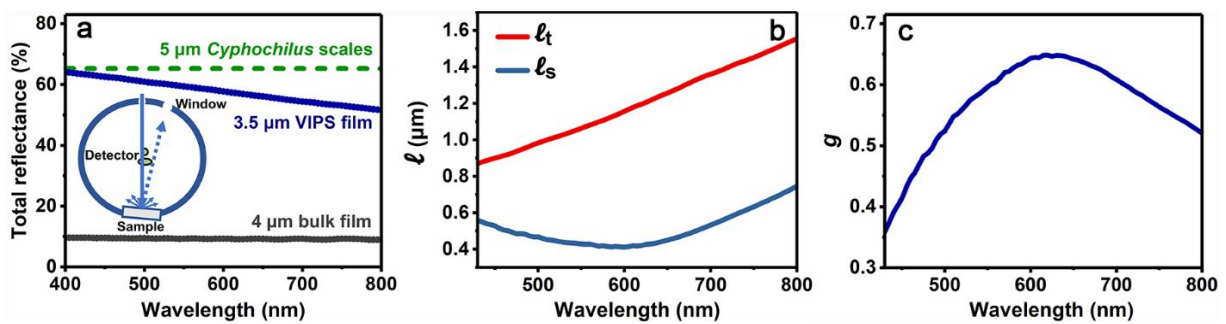

**Figure 2.** (a) Integrated reflectance spectrum for the 3.5 μm PS film at visible wavelengths, excluding specular reflections. The average reflection for a 5 μm *Cyphochilus* scale and the specular reflection for a transparent bulk PS film are also reported for comparison. (b) Transport (red line) and scattering (navy line) mean free paths and (c) scattering anisotropy *g* retrieved for the 3.5 μm PS film in the visible range.

To illustrate the formation mechanism of the bi-continuous network structure, we prepared different films varying the rotating speed of the spin-coater and the solution volume used under the same solution concentration of 25 wt% and the relative humidity of 95%. The resulting films (**Figures 3**) show that the bi-continuous topology is progressively lost as the film thickness exceeds ~10 μm, above which it is superseded by a more isotropic cellular-type structure. Additional structural changes associated with thicker samples include an increase of the air phase volume fraction (see Experimental Section), a reduced degree of structural anisotropy and the formation of a dense skin layer on the external surface (see **Figure S2**), as opposed to the porous surface that is characteristic of bi-continuous structures obtained by VIPS method.[42-45] For intermediate thickness, however, the two structures coexist as can be seen in Figure 3c for the 10.3 μm film, where the bi-continuous structure is still present near the surface.

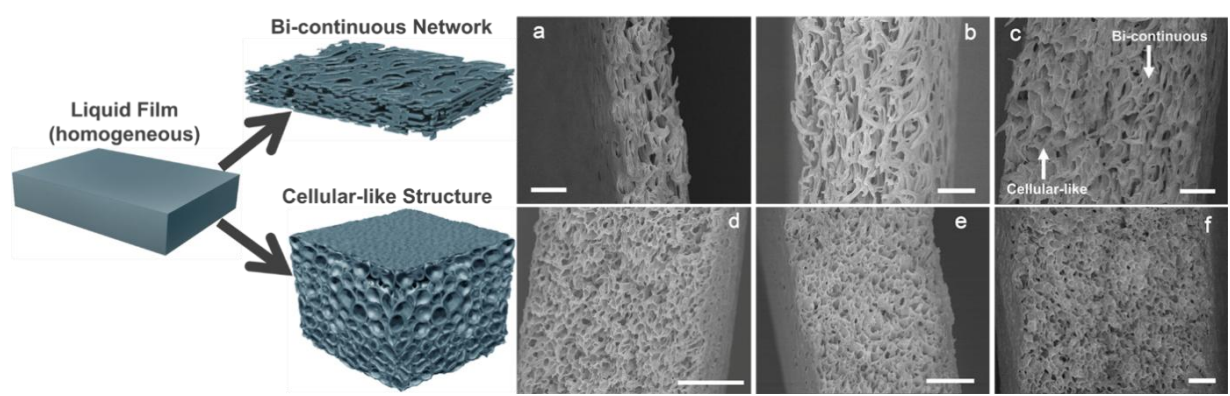

**Figure 3.** Sketch and SEM images of the cross-sections of films prepared by VIPS method. Average film thickness is 3.5, 6.0, 10.3, 18.0, 28.5 and 84 μm for panels a-f, respectively. All films have the same orientation, with the external surface on the right and the bottom surface on the left side. Scale bars: 2 μm in panels a-c and 10 μm in panels d-f.

As schematically shown in Figure 3, the occurrence of the bi-continuous network can be explained using mass-transfer-diffusion theory.[44,45] In general, there are two mechanisms for VIPS: nucleation-growth (NG) and spinodal decomposition (SD).[46] When the polymer solution is at the metastable state, phase separation occurs via the NG mechanism, leading to a cellular-like porous structure. On the other hand, SD represents the dominant phase separation mechanism when the system is at the unstable state, leading to the bi-continuous network structure.[47,48] Due to the affinity between water and the DMF solvent, water vapor diffuses into the liquid film during the spin-coating and solvent evaporation process. This results in a reduced polymer solubility which brings the solution into the metastable state. If enough water diffuses into the solution in a short time, the phase separation degree of the polymer solution is aggravated and the solution transits quickly from the metastable to the unstable state. Thin films under high relative humidity undergo a rapid water content increase, leading to the formation of the bi-continuous structure. Conversely, at higher thicknesses and/or lower relative humidity (see also control experiments at reduced relative humidity in Supporting information, **Figure S3**), the system will remain in the metastable state long enough for the NG process to take place, resulting in a typical cellular-like structure. Another important factor influencing the phase separation is polymer concentration. In fact, polymer concentration determines the so-called critical residence time of the system in the metastable state due to the entanglement of polymer chains.[49] A lower concentration is associated with a shorter critical residence time in the metastable region, which allows the NG process to complete rapidly, hindering the subsequent unstable state to form a bi-continuous structure. As shown in the control experiments (**Figure S4**), thin films (6-7 μm thick) prepared at 95% RH from 15-20 wt% PS solutions also show a cellular-like morphology. For these reasons, high polymer concentration, high relative humidity and thin solution thickness are required to form the more efficiently scattering bi-continuous structure akin to that of the *Cyphochilus* beetle.

The superior optical properties of the bi-continuous topology as opposed to the cellular-like one are confirmed by our optical characterization. In particular, **Figure 4**a shows the total reflectance and scattering efficiency at a wavelength of 500 nm attained by the different films shown in Figure 3, resulting in a comparable transport mean free path to that measured for the *Cyphochilus* beetle[3] as long as the film thickness remains below 10 μm, i.e., before the morphology transition to the isotropic, cellular-like structure. The complete reflectance spectra are reported in **Figure S5** and in **Figure S6** for samples prepared under different conditions, confirming the enhanced scattering efficiency of samples characterized by the denser, anisotropic bi-continuous structure. A comparison between the optical performance of the VIPS of PS films and other artificial bright-white materials made with low refractive index scattering elements is presented in **Table 1**.

**Table 1.** Average transport mean free path at a wavelength of 500 nm for typical artificial low-*n* highly scattering media.

| Material | $\ell_t$ (μm) |
|---|---|
| **PS film (this work)** | **0.98 ± 0.03** |
| PMMA film by NIPS[26] | ~1 |
| Photonic glass[3] | 2.9 |
| SC-$CO_2$ foaming surface[24] | 3.5-4 |
| CNC paper[14] | ~4 |
| P(VdF-HFP)$_{HP}$ PDRC coatings[5] | ~6 |
| Syringe filter[3] | 6 |

The transition from the anisotropic bi-continuous structure to the cellular-like structure is also apparent by the different spectral dependence of the retrieved optical properties. In particular, the structural transition is accompanied by both a different power-law dependence of the reduced scattering coefficient (defined as $\mu_s'=1/\ell_t$, Figure 4b) and a change of the scattering mean free path $\ell_s$ (Figure 4c). As regards the asymmetry parameter *g*, its spectral dependence in bi-continuous films exhibits a peak of increased forward scattering at visible frequencies, as compared to the flatter response for cellular-like structures (Figure 4d). In this case, we retrieved the asymmetry factor only for the 18.0 μm thick film, since this is the thickest sample that still exhibits a detectable ballistic transmission signal. A red-shift of the parameter *g* peak is also observed, which is in qualitative agreement with a simple model based on light scattering from infinite cylinders with a gaussian radii distribution, suggesting that thicker films are characterized by larger and more polydisperse scattering elements as confirmed by SEM images of the same samples (see Supporting Information, **Figure S7**). Considered together, the increased structural isotropy, polydispersity and filament size observed for thicker PS films account for their decreased overall scattering efficiency.[13,17,19] SEM images show that an optimal filament size is obtained in the ultra-thin films, close to that of (244±51) nm measured for the *Cyphochilus* beetle and well within the ideal 200-300 nm range determined theoretically (**Figure S8**).[17]

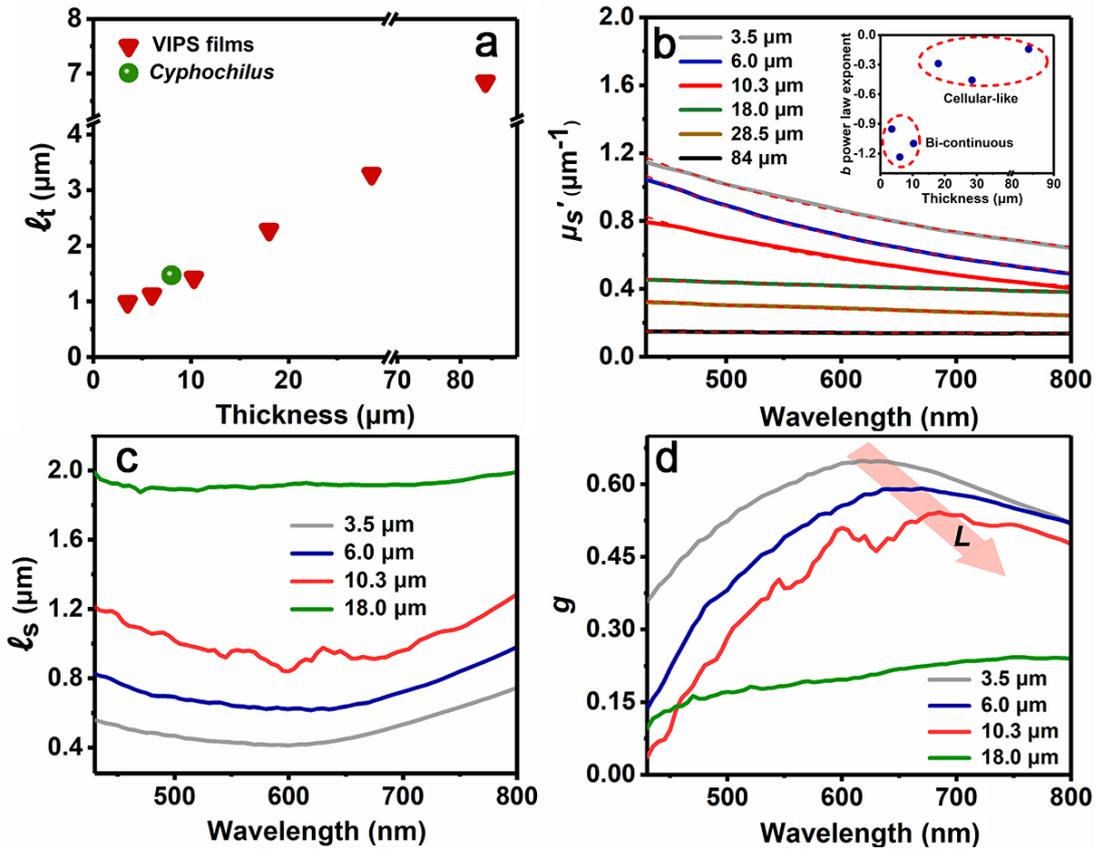

**Figure 4.** (a) Transport mean free path at 500 nm for different PS film thicknesses and for the *Cyphochilus* beetle.[3] (b) Reduced scattering coefficient and power-law fitting with the function $a\, x^b$. Fitted exponents are plotted in the inset, roughly forming two groups with about $b=0$ and $b=-1$ relative to the scattering behavior of cellular-like and bi-continuous structures. (c) Spectral dependence of scattering mean free path retrieved for thin PS films. (d) Spectral dependence of the asymmetry factor retrieved for thin films, based on the ballistic component

of transmitted intensity. A red shift ($L$) of ~60 nm in bi-continuous samples is observed, with no obvious peak in the cellular-like 18.0 μm thick film.

Besides their utility as diffusing layers, the highly porous structure of these thin films lends itself to other interesting applications. In particular, their morphology allows liquids, vapors and gases to percolate through the structure, which can cause dramatic changes in the scattering properties enabling optical sensing applications.[26,50-52] Based on the excellent water permeability of the bi-continuous network structure[43], herein we demonstrate a potential application of the biomimetic film as a real-time optical moisture sensor. The surface wettability of the ultra-thin film was changed from hydrophobic (static contact angle 81°) to super-hydrophilic (static contact angle 0°) by cold air plasma treating (**Figure 5**a, 5b and **Figure S9**).[53,54] When immersed in water, the treated film immediately switches from a white opaque to a transparent appearance due to the reduced refractive index contrast caused by water infiltration (Figure 5b). As an exemplary application exploiting this effect, we describe the use of the bright-white super-hydrophilic film as an optical moisture sensor for real-time monitoring of human exhalation. As shown in Figure 5d-e using the 3.5 μm sample, different exhalation rates and durations give rise to different optical signals according to the rapid change in reflectivity of the sample. This example illustrates the potential use of ultra-thin, super-hydrophilic bright-white films as fast responding sensing elements for human breath.

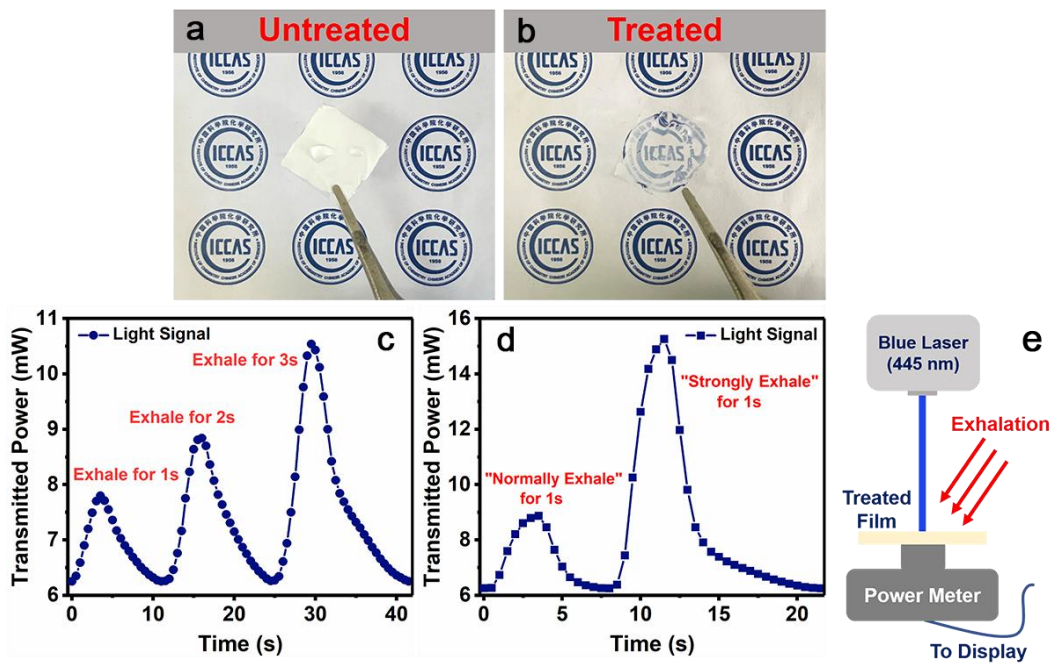

**Figure 5.** Appearance of (a) original and (b) cold plasma treated ultra-thin bi-continuous films after being dipped into water. (c, d) Different exhalation durations and rates measured as a change of detected power. (e) Sketch of exhalation sensing setup.

In conclusion, inspired by the scales of the *Cyphochilus* beetle, we developed a simple water-vapor induced phase separation method to prepare a series of ultra-thin bright-white PS films with optical performances comparable or superior to those of the white beetle and previously reported artificial materials, with the additional advantages given by free-standing stability, flexibility, and activatable super-hydrophilicity which make them compatible with a wide array of optical devices and applications. Due to the applicability of the phase-separation method to many polymers and the convenience of the coating techniques, we envision that diverse and large-area ultra-thin bright-white polymeric films might establish as a competitive alternative to nanoparticle-based coatings. By performing a broadband optical characterization of the scattering properties based on the rigorous solution of the radiative transfer model, we revealed how structural transitions are associated to spectral signatures at the single and multiple scattering level. Indeed, up to date, the description of optical properties in bright-white materials has been largely limited to the framework of diffusive approximation, disregarding the role of the scattering asymmetry associated to their network structure. Especially for the case of ultra-thin scattering materials, the use of more accurate radiative transport models providing information down to the single scattering level can provide further insight needed to compare the increasing amount of advanced optical materials, including the *Cyphochilus* beetle itself, and understand if they rely on similar scattering mechanisms or if they are characterized by a different interplay between scattering parameters. On a related note, even though spinodal decomposition of biomacromolecules such as chitin is currently supposed to be the driving mechanism behind the formation of bi-continuous network materials in the biological world, conclusive evidence on this matter is still lacking.[55] It is worth noting that our VIPS method requires only mild conditions for non-solvent induced spinodal decomposition as potentially found in most organisms, and results in samples that are similar to the white *Chyphochilus* scales both in terms of morphology and scattering performance. In this respect, optical characterization at the single-scattering level of self-assembled polymeric structures might provide a valuable tool to shed light on the growth process of white beetle scales and help characterizing the different morphological pathways that are possibly playing a role in their formation.

**Experimental Section**

*Materials*: Polystyrene ($M_w$ = 250 kDa, purchased from Acros), *N,N'*-dimethylformamide (DMF, analytical grade, purchased from Beijing Chemical Works), potassium carbonate ($K_2CO_3$, 99.5%, guaranteed reagent, purchased from Innochem) and sodium chloride (NaCl, analytical grade, purchased from Beijing Chemical Works) were used as received.

*Film preparation process*: Granular PS was dissolved in DMF under mechanical agitation at 80 °C for 24 hours. Granular PS was dissolved in DMF under mechanical agitation at 80 °C for 24 hours. A spin-coater (KW-4A, Setcas, China) was covered by a hood. A culture dish containing water or aqueous solution of salt was placed in the hood to control the relative humidity, which was monitored by a hygrometer (608-H2, error of ±2%, Testo AG, Germany). A relative humidity (RH) of 95±2% was obtained by the natural evaporation of water in the sealed spin-coater at room temperature. RHs of 43% and 71% were maintained by using the supersaturation solutions of $K_2CO_3$ and NaCl, respectively.[56] Polymer solution was dropped via a syringe and spin coated onto a cover slip ($2\times2$ cm$^2$). A liquid film formed on the slip in less than 1 second. Then the spin-coater was turned off, and the liquid film was kept in the humid atmosphere to evaporate the solvent. By controlling the injection volume of polymer solution and the spinning speed, the thickness of the films prepared could be adjusted from a few microns to tens of microns. Considering significant thickness fluctuations and poor mechanical stability due to excessive spinning speed or too little polymer solution covering the substrate, the limit of film thickness is controlled above 3 μm. Specifically, films of 3.5, 6.0, 10.3, 18.0, 28.5 and 84 μm were prepared by using approximately 0.01 mL at 9000 r min$^{-1}$, 0.01 mL at 8000 r min$^{-1}$, 0.02 mL at 8000 r min$^{-1}$, 0.05 mL at 7500 r min$^{-1}$, 0.08 mL at 5000 r min$^{-1}$ and 0.25 mL at 2500 r min$^{-1}$, respectively. Following solvent evaporation, the obtained films were taken out and dried in a vacuum oven at 70 ℃ overnight. The photo of the setup is shown in **Figure S10** of Supporting Information.

*Total reflectance measurement*: Total reflectance was measured by a UV/VIS/NIR spectrometer (Lambda 950, Perkin Elmer, USA) with a 150 mm integrating sphere. A light beam of about $0.3\times1$ cm$^2$ was incident at 8 degrees and a window at −8 degrees was opened to eliminate the specular reflection (except for measurement of transparent bulk PS film), which was about 2% in our case, as shown in the inset of Figure 2a. A common beam depolarizer was used to obtain unpolarized incident light. A standard white diffuser (SRS-99-020, Labsphere, USA) was used for the calibration before measurement.

*Angular-resolved measurement*: A halogen lamp (HL2000, Ideaoptics, China) was fixed along the direction orthogonal to the film surface in a goniometer (R1, Ideaoptics, China). The detector (NOVA-EX, Ideaoptics, China) connected with goniometer by a 600 μm core optic fiber (FIB-M-600-NIR, Ideaoptics, China) recorded scattering light signal from 300 to 1000 nm wavelength at different angles subtending a solid angle 3.9 °. The light source was fixed at 0 ° and the detector was rotated from 90 ° to 270 ° in the transmittance mode. Ballistic light was estimated as the ratio between the diffused light intensity extrapolated at 180 ° and the actual intensity collected at 180 °. Integration time was adjusted from 1000 to 4000 ms depending on the film thickness. Agular measurements are taken in steps of 5 °.

*Morphological characterization*: The morphology of the films was observed by SEM (JSM-7500F, JEOL, Japan) at an accelerating voltage of 5 kV. The films were coated with platinum by a sputter coater (JEC-3000FC, JEOL, Japan) before investigation. Film thickness was measured using the cross-section SEM image and an average data of 20 different points was reported. The thickness of samples is 3.5±0.1, 6.0±0.1, 10.3±0.1, 18.0±0.3, 28.5±0.5, 84±2 μm, respectively.

*Volume fraction of microstructure*: We cut samples in rectangular films and estimated the apparent density as $\rho = m/V$ considering the average film thickness. The volume fraction was calculated using a nominal density of 1.04 g/cm$^3$ for amorphous PS.[28] We obtained volume fractions of 0.59, 0.58, 0.48, 0.41, 0.41 and 0.40 for films of 3.5, 6.0, 10.3, 18.0, 28.5 and 84 μm thick, respectively.

*Static contact angle test*: Static contact angle (CA, $\theta$) was measured using a drop shape analysis instrument (KRÜSS DSA 100, Germany) at ambient temperature via a sessile drop method. A water droplet of 5 μL was employed on the untreated sample and 1 μL was employed on the treated sample and each contact angle value was an average of three measurements on different positions of the surface.

*Exhalation monitoring test*: A 445 nm blue laser diode source was used delivering an optical power of 117 mW as measured by an optical power meter (Model 843-R, Newport, USA) equipped with a 919P thermopile sensor without the scattering sample. All tests were carried out in ambient environment with a RH of 35±2%.

**Supporting Information**
Supporting Information is available from the Wiley Online Library or from the author.

**Acknowledgements**
W.Z. and L.P. contributed equally to this work. This work was financially supported by the National Nature Science Foundation of China (No.51673203, 51522308) and the Chinese Academy of Sciences (No. QYZDB-SSW-SLH025). W.Z., N.Z. and J.X. thank Prof. Jian Zi, Prof. Lei Shi and Dr. Yujie Bai at Fudan University for fruitful discussions; Ideaoptics company (Shanghai, China) for help in angular-resolved measurement; Dr. Zijian Xu and Dr. Xiangzhi Zhang at Shanghai Synchrotron Radiation Facility for help in morphological analysis. L.P. and D.S.W. acknowledge funding from Ente Cassa di Risparmio Firenze, Prog. No. 2015-0781 and Laserlab-Europe, H2020 EC-GA (654148).

Received: ((will be filled in by the editorial staff))
Revised: ((will be filled in by the editorial staff))
Published online: ((will be filled in by the editorial staff))

References

[1] P. Vukusic, B. Hallam, J. Noyes, *Science* **2007**, *315*, 348.

[2] D. S. Wiersma, *Nat. Photonics* **2013**, *7*, 188.

[3] M. Burresi, L. Cortese, L. Pattelli, M. Kolle, P. Vukusic, D. S. Wiersma, U. Steiner, S. Vignolini, *Sci. Rep.* **2014**, *4*, 6075.

[4] P.-C. Hsu, A. Y. Song, P. B. Catrysse, C. Liu, Y. Peng, J. Xie, S. Fan, Y. Cui, *Science* **2016**, *353*, 1019.

[5] J. Mandal, Y. Fu, A. Overvig, M. Jia, K. Sun, N. Shi, H. Zhou, X. Xiao, N. Yu, Y. Yang, *Science* **2018**, eaat9513.

[6] S. H. Choi, S.-W. Kim, Z. Ku, M. A. Visbal-Onufrak, S.-R. Kim, K.-H. Choi, H. Ko, W. Choi, A. M. Urbas, T.-W. Goo, Y. L. Kim, *Nat. Commun.* **2018**, *9*, 452.

[7] G. R. R. Bell, L. M. Mäthger, M. Gao, S. L. Senft, A. M. Kuzirian, G. W. Kattawar, R. T. Hanlon, *Adv. Mater.* **2014**, *26*, 4352.

[8] H. Luo, J. K. Kim, E. F. Schubert, J. Cho, C. Sone, Y. Park, *Appl. Phys. Lett.* **2005**, *86*, 243505.

[9] T. W. Koh, J. A. Spechler, K. M. Lee, C. B. Arnold, B. P. Rand, *ACS photonics* **2015**, *2*, 1366.

[10] B. Pyo, C. W. Joo, H. S. Kim, B.-H. Kwon, J.-I. Lee, M. C. Suh, *Nanoscale* **2016**, *8*, 8575.

[11] J. Xue, Y. Gu, Q. Shan, Y. Zou, J. Song, L. Xu, Y. Dong, J. Li, H. Zeng, *Angew. Chem. Int. Edit.* **2017**, *129*, 5316.

[12] J. D. Forster, H. Noh, S. F. Liew, V. Saranathan, C. F. Schreck, L. Yang, J.-G. Park, R. O. Prum, S. G. J. Mochrie, C. S. O'Hern, H. Cao, E. R. Dufresne, *Adv. Mater.* **2010**, *22*, 2939.

[13] L. Cortese, L. Pattelli, F. Utel, S. Vignolini, M. Burresi, D. S. Wiersma, *Adv. Opt. Mater.* **2015**, *3*, 1337.

[14] S. Caixeiro, M. Peruzzo, O. D. Onelli, S. Vignolini, R. Sapienza, *ACS Appl. Mater. Interfaces* **2017**, *9*, 7885.

[15] G. Buxbaum, *Industrial Inorganic Pigments*. Wiley, **2008**.

[16] C. Åkerlind, H. Arwin, T. Hallberg, J. Landin, J. Gustafsson, H. Kariis, K. Järrendahl, *Appl. Optics* **2015**, *54*, 6037.

[17] S. M. Luke, B. T. Hallam, P. Vukusic, *Appl. Optics* **2010**, *49*, 4246.

[18] H. L. Leertouwer, B. D. Wilts, D. G. Stavenga, *Opt. Express* **2011**, *19*, 24061.

[19] B. D. Wilts, X. Sheng, M. Holler, A. Diaz, M. Guizar-Sicairos, J. Raabe, R. Hoppe, S. H. Liu, R. Langford, O. D. Onelli, D. Chen, S. Torquato, U. Steiner, C. G. Schroer, S. Vignolini, A. Sepe, *Adv. Mater.* **2018**, *30*, 1702057.

[20] L. Pattelli, A. Egel, U. Lemmer, D. S. Wiersma, *Optica* **2018**, *5*, 1037.

[21] T. G. Smijs, S. Pavel, *Nanotechnol. Sci. Appl.* **2011**, *4*, 95.

[22] H. Shi, R. Magaye, V. Castranova, J. Zhao, *Part. Fibre Toxicol.* **2013**, *10*, 15.

[23] M. J. A. Ruszala, N. A. Rowson, L. M. Grover, R. A. Choudhery, *Int. J. Chem. Eng. Appl.* **2015**, *6*, 331.

[24] J. Syurik, R. H. Siddique, A. Dollmann, G. Gomard, M. Schneider, M. Wordull, G. Wiegand, H. Hölscher, *Sci. Rep.* **2017**, *7*, 46637.

[25] F. Zeighami, M. A. Tehran, *J. Ind. Text.* **2016**, *46*, 495.

[26] J. Syurik, G. Jacucci, O. D. Onelli, H. Hölscher, S. Vignolini, *Adv. Funct. Mater.* **2018**, 1706901.

[27] S. R. Sellers, W. Man, S. Sahba, M. Florescu, *Nat. Commun.* **2017**, *8*, 14439.

[28] D. Schrader, in *Polymer Handbook*, Vol. 2 (Eds: J. Brandrup, E. H. Immergut, E. A. Grulke, A. Abe, D. R. Bloch), Wiley, New York, USA **1989**

[29] G. Mazzamuto, L. Pattelli, C. Toninelli, D. S. Wiersma, *New J. Phys.* **2016**, *18*, 023036.

[30] W. Wu, H. Haick, *Adv. Mater.* **2018**, 1705024.

[31] H. Jin, Y. S. Abu‐Raya, H. Haick, *Adv. Healthcare Mater.* **2017**, *6*, 1700024.

[32] G, Konvalin, H. Haick, *Acc. Chem. Res.* **2014**, *47*, 66.

[33] A. Ishimaru, *Wave Propagation and Scattering in Random Media*, Wiley, **1999**.

[34] R. Gómez-Medina, L. S. Froufe-Pérez, M. Yépez, F. Scheffold, M. Nieto-Vesperinas, J. J. Sáenz, *Phy. Rev. A* **2012**, *85*, 035802.

[35] B. X. Wang, C. Y. Zhao, *Phy. Rev. A* **2018**, *97*, 023836.

[36] D. J. Durian, *Phys. Rev. E* **1994**, *50*, 857.

[37] K. M. Yoo, R. R. Alfano, *Opt. Lett.* **1990**, *15*, 320.

[38] R. Elaloufi, R. Carminati, J. J. Greffet, *J. Opt. Soc. Am. A* **2004**, *21*, 1430.

[39] E. Alerstam, *Phys. Rev. E* **2014**, *89*, 063202.

[40] A. Fernandez-Oliveras, M. Rubiño, M. M. Perez, *J. Eur. Opt. Soc-Rapid* **2012**, *7*.

[41] I. A. Mjör, *Dentin and Dentinogenesis* **1984**, *1*, 1.

[42] Y. S. Su, C. Y. Kuo, D. M. Wang, J. Y. Lai, A. Deratani, C. Pochat, D. Bouyer, *J. Membrane Sci.* **2009**, *338*, 17.

[43] J. T. Tsai, Y. S. Su, D. M. Wang, J. L. Kuo, J. Y. Lai, A. Deratani, *J. Membrane Sci.* **2010**, *362*, 360.

[44] V. P. Khare, A. R. Greenberg, W. B. Krantz, *J. Membrane Sci.* **2005**, *258*, 140.

[45] Y. Yip, A. J. McHugh, *J. Membrane Sci.* **2006**, *271*, 163.

[46] C. Y. Kuo, H. A. Tsai, J. Y. Lai, D. M. Wang, A. Deratani, C. Pochat-Bohatier, H. Matsuyama, ICOM, Seoul, South Korea, **2005**.

[47] I. M. Lifshitz, V. V. Slyozov, *J. Phys. Chem. Sol.* **1961**, *19*, 35.

[48] J. W. Cahn, *J. Chem. Phys.* **1965**, *42*, 93.

[49] S. L. Su, D. M. Wang, J. Y. Lai, *J. Membrane Sci.* **2017**, *529*, 35.

[50] J. M. Corres, Y. R. Garcia, F. J. Arregui, I. R. Matias, *IEEE Sens. J.* **2011**, *11*, 2383.

[51] B. Ding, M. Wang, J. Yu, G. Sun, *Sensors*, **2009**, *9*, 1609.

[52] J. Cimadoro, L. Ribba, S. Ledesma, S. Goyanes, *Macromol. Mater. Eng.* **2018**, *303*, 1800237.

[53] R. W. Paynter, *Surf. Interface Anal.* **2000**, *29*, 56.

[54] T. Murakami, S. Kuroda, Z. Osawa, *J. Colloid Interf. Sci.* **1998**, *202*, 37.

[55] S. L. Burg, A. J. Parnell, *J. Phys.: Condens. Matter* **2018**, *30*, 413001.

[56] W. M. Haynes, *CRC Handbook of Chemistry and Physics*, CRC press, **2014**.